\documentclass[twocolumn,tighten,twocolappendix]{aastex63}

\received{March, 3, 2021}
\revised{June, 3, 2021}
\accepted{June, 5, 2021}
\submitjournal{ApJ}

\shorttitle{SMS Formation in Magnetized Clouds}
\shortauthors{Hirano, Machida, and Basu}

\newcommand{\zsun}{Z_\odot}

\newcommand{\msun}{M_\odot}

\newcommand{\nh}{n_{\rm H}}
\newcommand{\cc}{{\rm cm^{-3}}}

\newcommand{\bp}{\beta_{\rm p}}
\begin{document}

\title{Supermassive Star Formation in Magnetized Atomic-Cooling Gas Clouds:\\
Enhanced Accretion, Intermittent Fragmentation, and Continuous Mergers}

\correspondingauthor{Shingo Hirano}
\email{hirano@astron.s.u-tokyo.ac.jp}

\author[0000-0002-4317-767X]{Shingo Hirano}
\affiliation{Department of Earth and Planetary Sciences, Faculty of Science, Kyushu University, Fukuoka 819-0395, Japan}
\affiliation{Department of Astronomy, School of Science, University of Tokyo, Tokyo 113-0033, Japan}

\author[0000-0002-0963-0872]{Masahiro N. Machida}
\affiliation{Department of Earth and Planetary Sciences, Faculty of Science, Kyushu University, Fukuoka 819-0395, Japan}
\affiliation{Department of Physics and Astronomy, University of Western Ontario, London, ON N6A 3K7, Canada}

\author[0000-0003-0855-350X]{Shantanu Basu}
\affiliation{Department of Physics and Astronomy, University of Western Ontario, London, ON N6A 3K7, Canada}

\begin{abstract}
The origin of supermassive black holes (with $\gtrsim\!10^9\,M_{\odot}$) in the early universe (redshift $z \sim 7$) remains poorly understood.
Gravitational collapse of a massive primordial gas cloud is a promising initial process, but theoretical studies have difficulty growing the black hole fast enough. 
We focus on the magnetic effects on star formation that occurs in an atomic-cooling gas cloud.
Using a set of three-dimensional magnetohydrodynamic (MHD) simulations, we investigate the star formation process in the magnetized atomic-cooling gas cloud with different initial magnetic field strengths.
Our simulations show that the primordial magnetic seed field can be quickly amplified during the early accretion phase after the first protostar formation.
The strong magnetic field efficiently extracts angular momentum from accreting gas and increases the accretion rate, which results in the high fragmentation rate in the gravitationally unstable disk region.
On the other hand, the coalescence rate of fragments is also enhanced by the angular momentum transfer due to the magnetic effects.
Almost all the fragments coalesce to the primary star, so the mass growth rate of the massive star increases due to the magnetic effects.
We conclude that the magnetic effects support the direct collapse scenario of supermassive star formation.
\end{abstract}


\section{Introduction} \label{sec:intro}

The formation mechanism of the supermassive black holes (SMBHs) with mass $\sim\!10^9\,\msun$ at an early epoch $z > 6$ (high-$z$ quasars) is one of the unsolved problems in the early universe \citep[see review by][]{woods19}\footnote{All $z > 5.7$ quasars currently known: \url{https://www.sarahbosman.co.uk/list_of_all_quasars}.}.
Each time the observations are updated \citep[e.g.,][]{yang20}, the demands on theoretical models become more stringent.

One of the popular formation channels of SMBHs is the direct collapse black hole (DCBH) scenario \citep[see review by][]{inayoshi19}.
In the pristine atomic-cooling (AC) halo, the gas component is cooled only by the atomic hydrogen and can gravitationally contract while remaining at a high temperature $\sim\!8000$\,K.
Because of the large Jeans mass due to the high temperature of the AC gas cloud, the cloud fragmentation is suppressed and the mass accretion rate onto the protostar increases.
Then the intermediate-mass black hole (IMBH) with $\sim\!10^5\,\msun$, a candidate seed of SMBHs, forms from the gravitationally-unstable AC halo.
If the protostellar mass exceeds the mass threshold ($\sim\,10^5\,\msun$) of the general relativistic instability, the star can collapse to form a massive black hole with a similar mass \citep{umeda16}.

To explain the observed number density of the SMBHs (a few Gpc$^{-3}$), the previous studies provided a number of conditions under which the AC gas cloud can be realized: H$_2$-dissociating ultraviolet radiation \citep[e.g.,][]{omukai01,agarwal12,latif13}, high-velocity collisions \citep{inayoshi15}, baryon-dark matter streaming velocities \citep{tanaka14,hirano17sv}, and dynamical heating due to the violent halo merger \citep{wise19}.
Recently, \cite{chon20} discussed the possibility of IMBH formation in a metal-enriched halo due to the {\it super competitive accretion} of a large number of fragments.

The magnetic field provides another physical process.
For the DCBH formation process in the early universe, the primordial magnetic field strength is too weak to affect the star formation process.
However, magnetic amplification mechanisms could enhance the ultraweak seed magnetic field due to a small-scale dynamo process during star-formation \citep[e.g.,][]{sur10,turk12,schober12,sharda20}. Subsequently, a strong magnetic field might affect the angular momentum transfer in the pristine star-forming region \citep{machida08,machida13}.

For the pristine magnetized AC gas cloud, the previous studies investigated the amplification of the magnetic field during the collapse phase \citep{latif13mag,grete19}.
However, the possible amplification of the magnetic field during the later accretion phase and how an amplified magnetic field contributes to protostar growth have not yet been studied.

In this study, we focus on the magnetic effects during the early accretion phase in the magnetized AC gas cloud for up to the first $1000$\,yr after the first protostar formation.
We perform a set of MHD simulations with different initial magnetic field strength, rotation rate, and metallicity of the initial AC gas cloud.
We investigate the impact of magnetic field, and whether or not it promotes massive star formation.

\begin{figure}[t]
\begin{center}
\includegraphics[width=1.0\columnwidth]{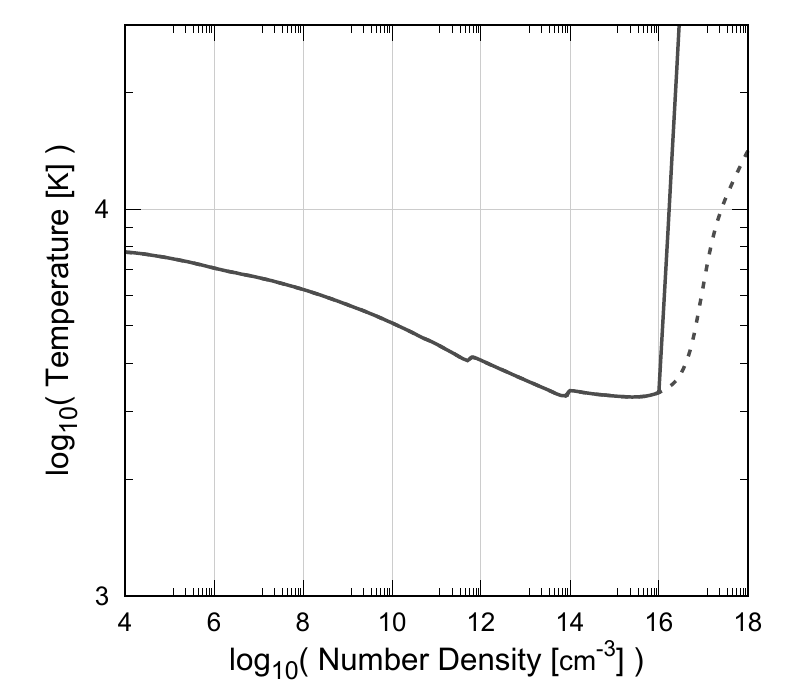}
\end{center}
\caption{
Thermal evolution made from tables of the equation of state (EOS) for a metallicity with $Z = 0\,\zsun$ (or $Z = 10^{-6}\,\zsun$ labelled in figure~6 of \citealt{omukai08}).
The adopted EOS in this study (solid lines) are variants of the theoretically constructed EOS based on the chemical reaction simulations \citep[dashed;][]{omukai08} using the stiff EOS technique with a threshold density $n_{\rm ad} = 10^{16}\,\cc$ (for details see \citealt{machida15}).
}
\label{f1}
\end{figure}

\section{Initial Condition and Numerical Settings} \label{sec:methods}

We solve the MHD equations as shown in equations (1)-(4) of \cite{machida13}.
We adopt EOS tables for a metallicity with $Z = 10^{-6}\,\zsun$ based on a chemical reaction simulation \citep{omukai08} that is the same as used by \citet{chon20}.
To accelerate the time evolution, we adopt a stiff EOS technique.
We set a threshold density $n_{\rm ad} = 10^{16}\,\cc$, which reproduces dense cores whose radius is consistent with the approximate mass--radius relation of a rapidly accreting protostar with $R \simeq 12\, (M/100\,\msun)^{1/2}\,{\rm au}$, as from equation (11) of \cite{hosokawa12} and figures~3, A1, and A2 of \cite{machida15}.
Figure~\ref{f1} shows the resultant EOS tables.
The thermal evolution with $Z = 10^{-6}\,\zsun$ is the same with the pristine chemistry of the zero-metallicity gas cloud.
Note that \citet{omukai08} clearly stated that the thermal evolution follows the metal-free (i.e., $Z = 0$) track when $Z \le 10^{-6}\,\zsun$.
Then we define models with the EOS with $Z = 10^{-6}\,\zsun$ as the pristine case ($0\,\zsun$) in this study.

The initial cloud has a Bonnor-Ebert (BE) density profile $\rho_{\rm BE}(r)$ with the central density $n_{\rm c,0} = 10^4\,\cc$ and temperature $T_0 = 7700$\,K. 
To promote the cloud contraction, we enhance the density by a factor $f$, and adopt the density profile as $\rho(r) = f \rho_{\rm BE}(r)$ for the initial cloud.
With $f = 1.2$, the mass and size of the initial cloud are $M_{\rm cl} = 1.82\times10^{6}\,\msun$ and $R_{\rm cl} = 2.09\times10^6$\,au = $10.1$\,pc, respectively.
With these settings, the ratio of thermal energy to the gravitational energy of the initial cloud is $\alpha_0 = 0.7$.

\begin{deluxetable}{llllrr}[t]
\tablecaption{Parameters of Initial Clouds and Calculation Results}
\tablecolumns{12}
\tablenum{1}
\tablewidth{0pt}
\tablehead{
  \colhead{1} &
  \colhead{2} &
  \colhead{3} &
  \colhead{4} &
  \colhead{5} &
  \colhead{6} \\
  \colhead{Model} &
  \colhead{$\beta_{\rm rot,0}$} &
  \colhead{$B_{z,0}$} &
  \colhead{$\beta_{\rm p,0}$} &
  \colhead{$t_{\rm col}$} &
  \colhead{$t_{\rm end}$} \\
  \colhead{} &
  \colhead{} &
  \colhead{(G)} &
  \colhead{} &
  \colhead{(yr)} &
  \colhead{(yr)}
}
\startdata
M100 & $10^{-1}$ & 0          & --                 & 2339175 &  579 \\
M112 &           & $10^{-12}$ & $2.7\times10^{17}$ & +60 &  469 \\
M110 &           & $10^{-10}$ & $2.7\times10^{13}$ & +469 &  422 \\
M108 &           & $10^{-8}$  & $2.7\times10^{9}$  & +633 &  374 \\
M106 &           & $10^{-6}$  & $2.7\times10^{5}$  & +602 &  271 \\
M105 &           & $10^{-5}$  & $2.7\times10^{3}$  & +583 &  158 \\
\hline
M200 & $10^{-2}$ & 0          & --                 & 1209600 &  529 \\
M220 &           & $10^{-20}$ & $2.7\times10^{33}$ & -105 &  475 \\
M215 &           & $10^{-15}$ & $2.7\times10^{23}$ & -46 &  537 \\
M212 &           & $10^{-12}$ & $2.7\times10^{17}$ & -98 &  708 \\
M210 &           & $10^{-10}$ & $2.7\times10^{13}$ & -72 &  584 \\
M208 &           & $10^{-8}$  & $2.7\times10^{9}$  & -108 &  365 \\
M206 &           & $10^{-6}$  & $2.7\times10^{5}$  & -98 &  195 \\
M205 &           & $10^{-5}$  & $2.7\times10^{3}$  & +840 &   96 \\
\hline
M300 & $10^{-3}$ & 0          & --                 & 1165868 &  773 \\
M312 &           & $10^{-12}$ & $2.7\times10^{17}$ & +33 &  728 \\
M310 &           & $10^{-10}$ & $2.7\times10^{13}$ & +24 & 1149 \\
M308 &           & $10^{-8}$  & $2.7\times10^{9}$  & -21 & 1103 \\
M306 &           & $10^{-6}$  & $2.7\times10^{5}$  & +2 &  213 \\
M305 &           & $10^{-5}$  & $2.7\times10^{3}$  & +907 &   54 \\
\enddata
\tablecomments{
Column (1): model name.
Columns (2) and (3): parameters $\beta_{\rm rot,0}$ (the ratio of the rotational energy to the gravitational energy) and $B_{z,0}$ (the initial magnetic field strength) of the initial cloud.
Column (4): plasma beta of the initial cloud.
Column (5): elapsed time at the first dense core formation, where for each value of $\beta_{\rm rot,0}$ the models with nonzero $B_{z,0}$ show the elapsed time relative to the corresponding non-magnetized model.
Column (6): elapsed time at the end of the simulation after the first dense core formation.
}
\label{t1}
\end{deluxetable}

A rigid rotation is adopted within the BE sphere, and a uniform magnetic field is imposed in the whole computational domain.
In Cartesian coordinates, the directions of the global magnetic field and the rotation axis are always parallel to the $z$-axis.
In addition, we impose a uniform metallicity for the whole domain.
Then we adopt two parameters: ratio of rotational energy to the gravitational energy of the initial cloud ($\beta_{\rm rot,0}$) and initial magnetic field strength ($B_{z,0}$).
Table~\ref{t1} summarizes the parameter combinations of the simulation models.
We define the model names by connecting the common logarithms of two model parameters.
We adopt the model M212 with ($\beta_{\rm rot,0}$, $B_{z,0}$/G) = ($10^{-2}$, $10^{-12}$) as the fiducial model of this study because it is the closest approximation to the environment of the AC gas cloud with the primordial chemistry and cosmological seed magnetic field strength \citep[$10^{-15}$\,G at $\nh = 1\,\cc$ as][]{xu08}.

We use our nested grid code, in which the rectangular grids of ($i$, $j$, $k$) = ($256$, $256$, $32$) are superimposed. 
The nested grid code is the same as used in \citet{machida15} except for the equation of state.
We use the index ``$l$'' to describe a grid level. 
The grid size $L(l)$ and cell width $h(l)$ of the $l$th grid are twice larger than those of ($l+1$)th grid (e.g., $L(l) = 2L(l+1)$ and $h(l) = 2h(l+1)$).
The spatial resolution is the same among all simulations.
The grid size and cell width of the $l = 1$ grid are $L(1) = 6.68\times10^7$\,au and $h(1) = 2.61\times10^5$\,au, respectively.
We set the maximum grid level as $l = 18$ and the finest grid has $L(18) = 510$\,au and $h(18) = 1.99$\,au, respectively.

During the simulations, we assume that the region where the gas density exceeds $n_{\rm ad}$ are the dense cores that host a protostar.
We analyze the number and masses of such dense cores to examine the fragmentation process.
We cease simulations when the total mass of the dense cores with $n > n_{\rm ad}$ exceeds $10^6\,\msun$ because of the critical mass of the general relativistic instability \citep{umeda16}.
However, in some models, the calculation timestep $dt$ becomes very short ($dt \ll 0.01$\,yr) before the total mass of the dense cores reaches $10^6\,\msun$, because the Alfv\'{e}n velocity becomes very high above the dense core and disk. 
For such models, we stop the calculation because we cannot follow the further time evolution of the system. 
Thus, we terminate the calculation either when the total core mass reaches $10^6\,\msun$ or when the calculation timestep becomes very short. 
We could calculate the time evolution of the system for about $100$--$1000$\,yr after the first protostar formation (see Table~\ref{t1}).

\section{Results} \label{sec:res}

We first show the result for the fiducial model (M212) and compare it with the corresponding non-magnetized model (M200) in \S\ref{sec:res:fid}.
Then, we discuss the dependence of the magnetic effects on the initial magnetic field strength ($B_{z,0}$) for models in the M2 series in \S\ref{sec:res:B}.
We also examine the dependence on the initial rotation of the gas cloud ($\beta_{\rm rot,0}$) among models in the M1, M2, and M3 series in \S\ref{sec:res:rot}.

\begin{figure*}[t]
\begin{center}
\includegraphics[width=1.0\linewidth]{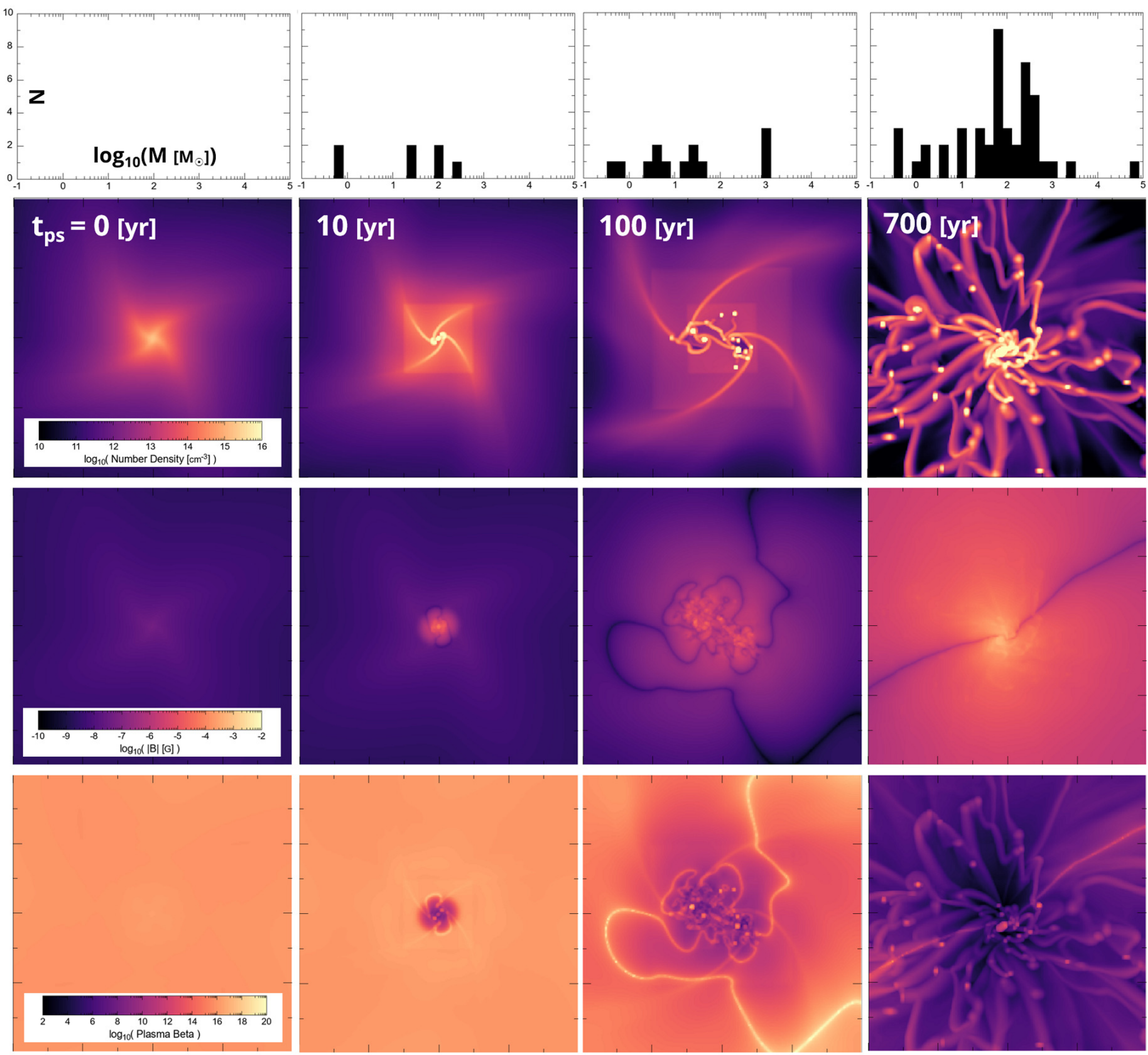}
\end{center}
\caption{
Time evolution for the fiducial model (M212 with $\beta_{\rm rot,0} = 10^{-2}$ and $B_{z,0} = 10^{-12}$\,G) at $t_{\rm ps} = 0$, $10$, $100$, and $700$\,yr after the first dense core formation (from left to right panels): mass spectrum of dense cores, number density, absolute magnetic field strength, and plasma beta on the $z = 0$ plane (from top to bottom).
The box size of 2D plot is $2000$\,au.
}
\label{f2}
\end{figure*}

\begin{figure}[t]
\begin{center}
\includegraphics[width=1.0\columnwidth]{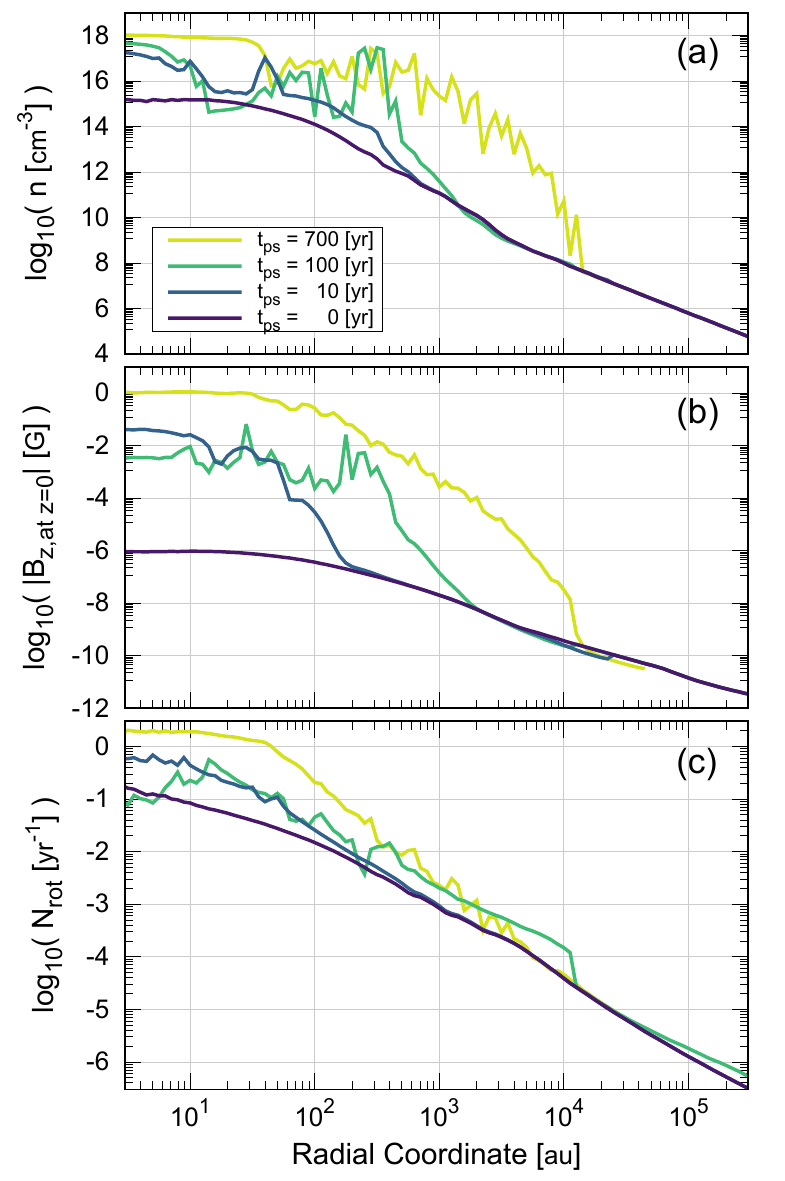}
\end{center}
\caption{
Radial profiles for the fiducial model (M212) at $t_{\rm ps} = 0$, $10$, $100$, and $700$\,yr (the same times in Figure~\ref{f2}). Panels: (a) number density, (b) absolute magnetic field strength on the $z = 0$ plane, and (c) rotational rate of the gas cloud ($N_{\rm rot} = 2 \pi \Omega$).
}
\label{f3}
\end{figure}

\begin{figure}[t]
\begin{center}
\includegraphics[width=1.0\columnwidth]{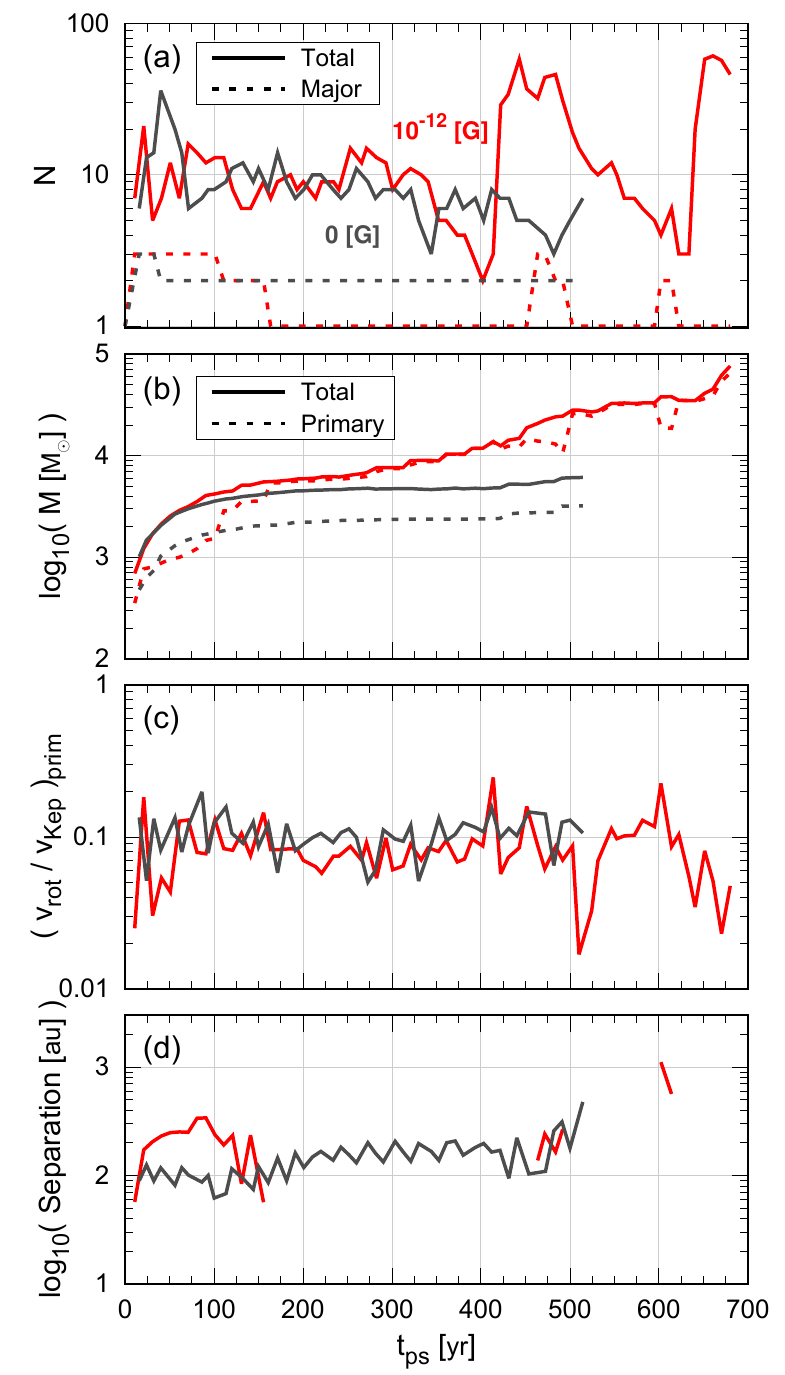}
\end{center}
\caption{
Time evolution of the core properties for the fiducial model (M212; red lines) and the zero magnetic field model (M200; black): (panel a) numbers of the total (solid) and major (dashed) cores, (b) total mass of cores (solid) and primary core mass (dashed), (c) ratio of the rotational velocity to the Keplerian velocity of the primary core, and (e) average separation among the major cores.
The primary core is the most massive core among all cores.
A major core is defined as a core with a mass greater than 10\% of the primary core mass.
}
\label{f4}
\end{figure}

\subsection{Fiducial model} \label{sec:res:fid}

First, we overview the simulation result for the fiducial model M212 with ($\beta_{\rm rot,0}$, $B_{z,0}$/G) = ($10^{-2}$, $10^{-12}$).
Figure~\ref{f2} shows the first $700$\,yr of evolution after the first dense core formation ($t_{\rm ps} = 0$)\footnote{It should be noted that, in Figure~\ref{f2}, the density in the finer inner grids appears to be higher than that in the coarser grids. This is an artefact of the plotting program. In Figure~\ref{f2} (and Figure~\ref{f8}), we used the gas density located closest to the $z=0$ plane, because there are no physical quantities exactly on the $z=0$ plane in our numerical code \citep[for reference, see also figure~3 of][]{matsumoto07}. Thus, the plotted quantities are slightly detached from the $z=0$ plane by $h(l)/2$, where $h(l)$ is the cell with of $l$-th grid.}.
We plot four panels from the top to bottom row: mass spectrum of the dense cores ($N_{\rm frag}$) and distributions of the gas number density ($n$), absolute magnetic field strength ($|B|$), and plasma beta ($\bp = P_{\rm thermal} / P_{\rm mag}$).
The massive gas cloud continuously fragments to a number of dense cores.
The dense cores intermittently coalesce into the most massive (primary) core.
A key result is that the magnetic field in the vicinity of the primary core is quickly amplified in the first $10$\,yr and the strong magnetic field region expands outward as the core formation proceeds.
As a result, the plasma beta declines in this expanding region, which means that the magnetic effects can affect the dynamics of the gas cloud.

Figure~\ref{f3} shows the radial profiles at the same epochs as in Figure~\ref{f2}.
After the primary dense core formation, the dense region expands outward due to the external gas accretion and the gravitationally unstable region fragments and forms a number of dense cores (Figure~\ref{f3}a).
As the dense region including a number of cores expands outwards, the amplified magnetic field region also expands outwards (Figure~\ref{f3}b).
The rapid amplification of the magnetic field in the vicinity of the dense core can be  interpreted as follows.
The region near the primary dense core rotates about once per yr (Figure~\ref{f3}c).
The magnetic field lines tied to the dense core are rapidly amplified as they are dragged by the high speed rotation of the core during the first $10$\,yr.
The rotation timescale is $<1$\,yr within $1$\,au, and a weak magnetic field passively traces the gas motion and is amplified. 
Thus, the exponential growth of the magnetic field naturally occurs in a short timescale, and the amplified field can move outward.

How does a strongly amplified magnetic field affect the massive star formation in the magnetized AC gas cloud?
Figure~\ref{f4} compares the time evolution of the dense core properties between the magnetized and non-magnetized models (M212 and M200).
The total numbers of dense cores are similar regardless of the magnetic effects (Figure~\ref{f4}a).
This is because  fragmentation and coalescence rates which are enhanced by the magnetic effects cancel each other out.
The angular momentum transfer due to the magnetic effects increases the mass accretion rate by weakening the rotation of the entire gas cloud.
The gravitationally unstable region gains more mass fuel and fragments to a number of dense cores.
On the other hand, the angular momentum transport due to the magnetic effects also promotes coalescence of dense cores.
The high mass accretion rate and efficient coalescence rate in the magnetized AC gas cloud speeds up the mass growth of the (primary) core (Figure~\ref{f4}b).

We also check the magnetic effects on the rotation velocity of the primary core and the averaged separation among the major cores with more than $0.1 M_{\rm frag,prim}$.
The rotation velocity of the star is the second most important parameter that determines its life and final fate.
Then there is a question whether the stellar rotation can be controlled by the amplified magnetic field \citep{hirano18}.
The simulations show that the rotation degree normalized by the Keplerian velocity is almost constant regardless of the presence of a magnetic field: $v_{\rm rot} \sim 0.1 v_{\rm Kep}$ (Figure~\ref{f4}c).
If the stars forming inside the core have the same degree of rotation, the evolutionary track and final fate of the first star would be little different from the no rotation case \citep{yoon12,chatzopoulos12}.
The averaged separations among the major cores reach a similar value, $\sim\!100$\,au (Figure~\ref{f4}d).
The angular momentum transport between the cores due to magnetic effects may have been canceled by the angular momentum brought in by the material falling in from the outside.

In the following subsections, we show the dependence of the magnetic effects on two model parameters: $B_{z,0}$ and $\beta_{\rm rot,0}$.

\subsection{Dependence on $B_{z,0}$} \label{sec:res:B}

First, we show the dependence on the initial magnetic field strength $B_{z,0}$.
The fiducial model, of which we have already presented results, initialized with $B_{z,0} = 10^{-12}$\,G, is consistent with the primordial magnetic field strength in the early universe.
We performed a set of simulations (M2 series) by changing $B_{z,0}$ from $10^{-20}$\,G to $10^{-5}$\,G (see Table~\ref{t1}).
The higher $B_{z,0}$ models can be used to study the cases with more efficient amplification of the magnetic field strength during the collapse phase.

\begin{figure*}[t]
\begin{center}
\includegraphics[width=1.0\linewidth]{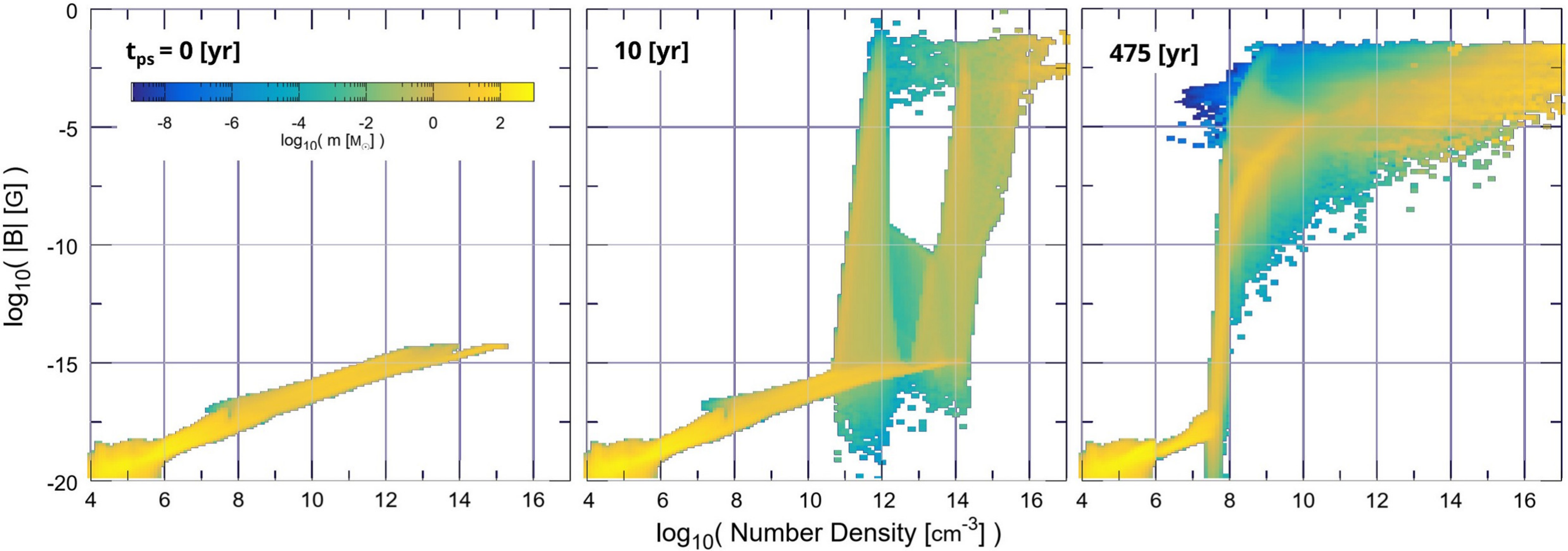}
\end{center}
\caption{
Phase diagrams of the absolute magnetic field strength for model M220 at $t_{\rm ps} = 0$, $10$, and $475$\,yr.
}
\label{f5}
\end{figure*}

\begin{figure*}[t]
\begin{center}
\includegraphics[width=0.9\linewidth]{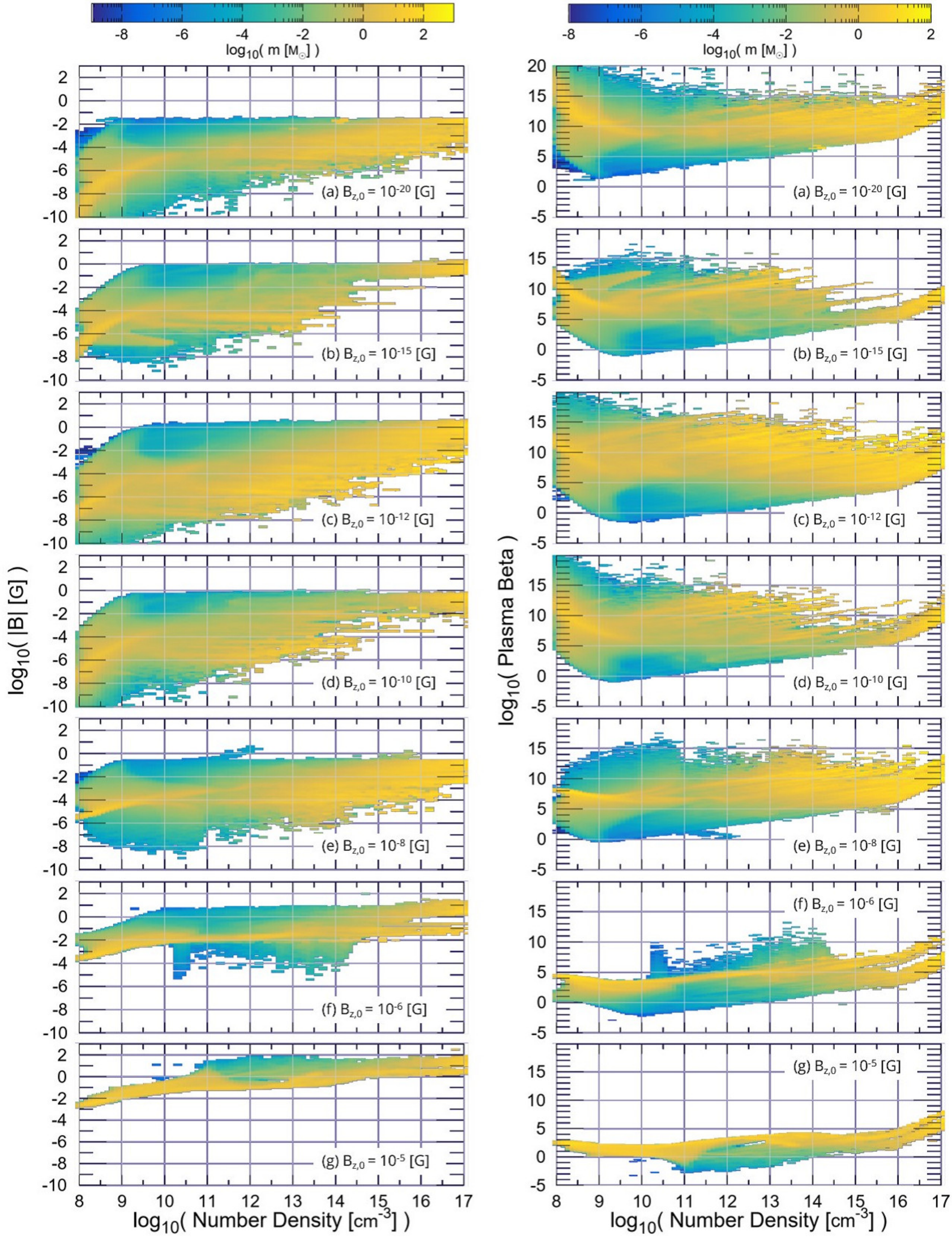}
\end{center}
\caption{
Phase diagrams of the absolute magnetic field strength (left panels) and plasma beta (right) at the final time for models with different initial magnetic field strengths $B_{z,0} = 10^{-20}$--$10^{-5}$\,G (M2 series: from top to bottom).
}
\label{f6}
\end{figure*}

\begin{figure}[t]
\begin{center}
\includegraphics[width=1.0\columnwidth]{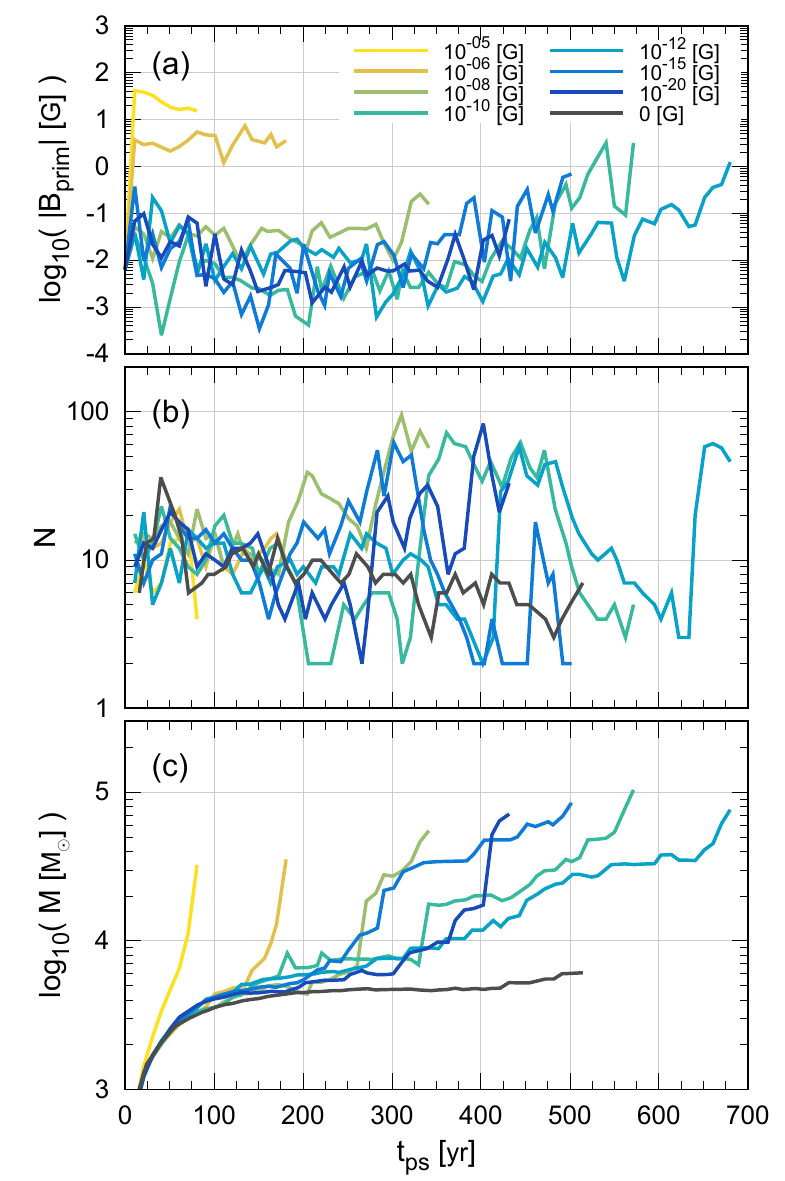}
\end{center}
\caption{
Time evolution of the core properties for models with different initial magnetic field strengths $B_{z,0} = 0$--$10^{-5}$\,G (M2 series). Panels: (a) absolute magnetic field strength of the primary core, (b) total number of fragments, and (c) total mass of cores.
}
\label{f7}
\end{figure}

Figure~\ref{f5} shows the time evolution of the magnetic field strength for the model initiated with the minimum $B_{z,0} = 10^{-20}$\,G (M220), which demonstrates the most extreme amplification.
When the first dense core forms at $t_{\rm ps} = 0$\,yr (the left panel), the magnetic field has been amplified only through flux freezing during the cloud compression, described by a power law $B \propto n_{\rm H}^{2/3}$.
In just the next $10$\,yr (the middle panel), the magnetic field strength is quickly amplified by more than $10$ orders of magnitude in the dense region where the dense cores form.
In the middle panel of Figure~\ref{f5}, the amplification of the magnetic field can be seen at two different densities of $\sim\!10^{12}\,\cc$ and $10^{14}-10^{16}\,\cc$, indicating that the magnetic field is amplified around both the low-density and high-density (or primary) cores at this epoch.
The strong magnetic field region grows and expands over time (the right panel).

Figure~\ref{f6} shows the mass distributions on the density--absolute magnetic field strength ($|B|$) plane and the density--plasma beta ($\bp$) plane at the final step of each model.
In the models with $10^{-15} \le B_{z,0}/{\rm G} \le 10^{-8}$, the magnetic field strength rapidly amplifies up to $1$\,G, regardless of the initial magnetic field strength (left panels).
As the magnetic field strength is amplified, the plasma beta, which is the ratio of the thermal pressure to the magnetic pressure, decreases rapidly (right panels).
This means that the magnetic effect cannot be ignored when considering the cloud evolution and fragmentation process.
In fact, the accretion rate to the dense cores increases as the magnetic field is amplified over time.
In the models with $B_{z,0} \ge 10^{-6}$\,G, the magnetic field strength exceeds $10^{-2}$\,G during the collapse phase and amplifies to more than $1$\,G.
In these models, the plasma beta is low even before the accretion phase, indicating that the magnetic effects are also important during the collapse phase.
We can see a bifurcation in the density--magnetic field strength and density--plasma beta planes, especially in panels of $B_{z,0} = 10^{-6}$ and $10^{-5}$\,G. 
The magnetic field strength around the core depends on the duration after the core formation.
Thus, the bifurcation means that there exist two (or three) massive cores formed at different epochs.

Figure~\ref{f7} shows the time evolution of the core's properties.
As shown above, all models show a similar trend regardless of the initial magnetic field strength, except for models with extremely strong initial magnetic field strength ($10^{-6}$--$10^{-5}$\,G).
The magnetic field strength around the primary core ($B_{\rm prim}$) is quickly amplified up to $10^{-2}$\,G just after the core formation and gradually increases during the accretion phase (Figure~\ref{f7}a).
The angular momentum transfer due to the magnetic effects enhances both fragmentation and coalescence rates.
The resultant number of cores varies between a few and a hundred due to fragmentation of the gravitationally unstable disk,  high mass accretion, and intermittent coalescence of cores (Figure~\ref{f7}b).
The mass growth of the primary core proceeds fastest for the model with the greatest initial magnetic field strength (Figure~\ref{f7}c).
For weaker $B_{z,0}$ the rapid mass growth also occurs, but at later times\footnote{For models with $B_{z,0} \le 10^{-12}$\,G, the mass growth of the primary core is faster in the order of M215, M220, and M212, and is not necessarily proportional to $B_{z,0}$. This may be due to a competition between angular momentum transport by magnetic effects and the suppression of mass accretion by magnetic effects (magnetic tension and magnetic pressure gradient forces).}.

\subsection{Dependence on $\beta_{\rm rot,0}$} \label{sec:res:rot}

\begin{figure*}[t]
\begin{center}
\includegraphics[width=1.0\linewidth]{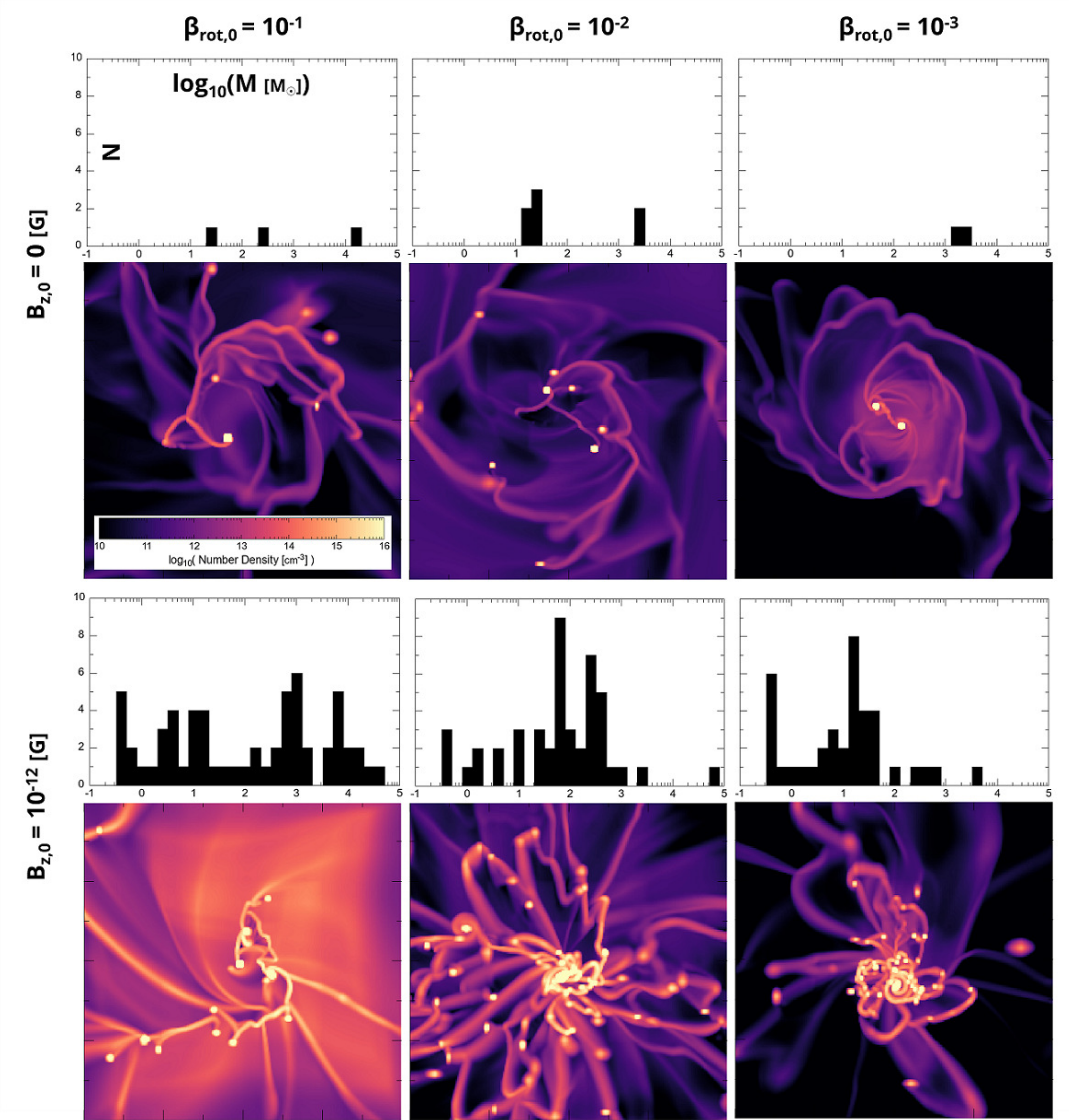}
\end{center}
\caption{
Core mass spectrum (top panel) and number density on the $z = 0$ plane (bottom) at the final step for models with two different initial magnetic field strengths $B_{z,0} = 0$ and $10^{-12}$\,G and three different initial rotation degrees $\beta_{\rm rot,0} = 10^{-1}$ (M100 and M112), $10^{-2}$ (M200 and M212), and $10^{-3}$ (M300 and M312).
The box size of the 2D plot is $2000$\,au.
}
\label{f8}
\end{figure*}

\begin{figure*}[t]
\begin{center}
\begin{tabular}{cc}
\includegraphics[width=1.0\columnwidth]{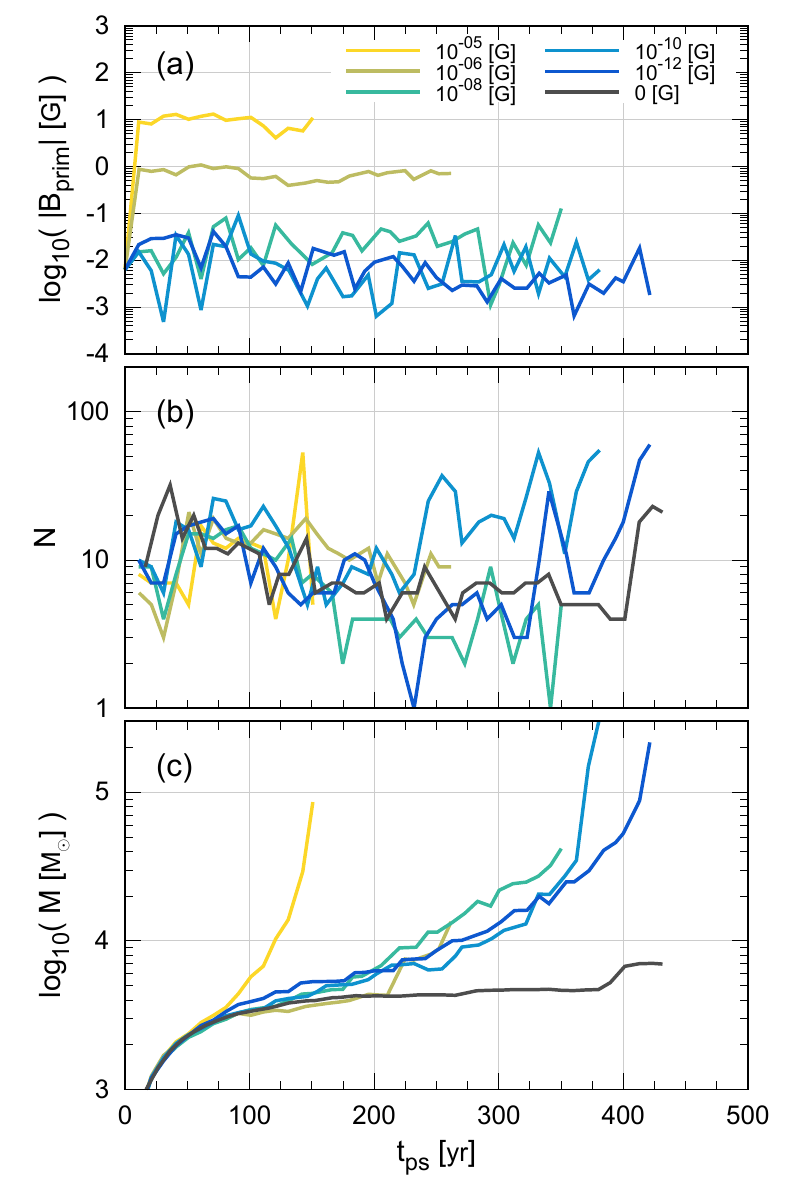}&
\includegraphics[width=1.0\columnwidth]{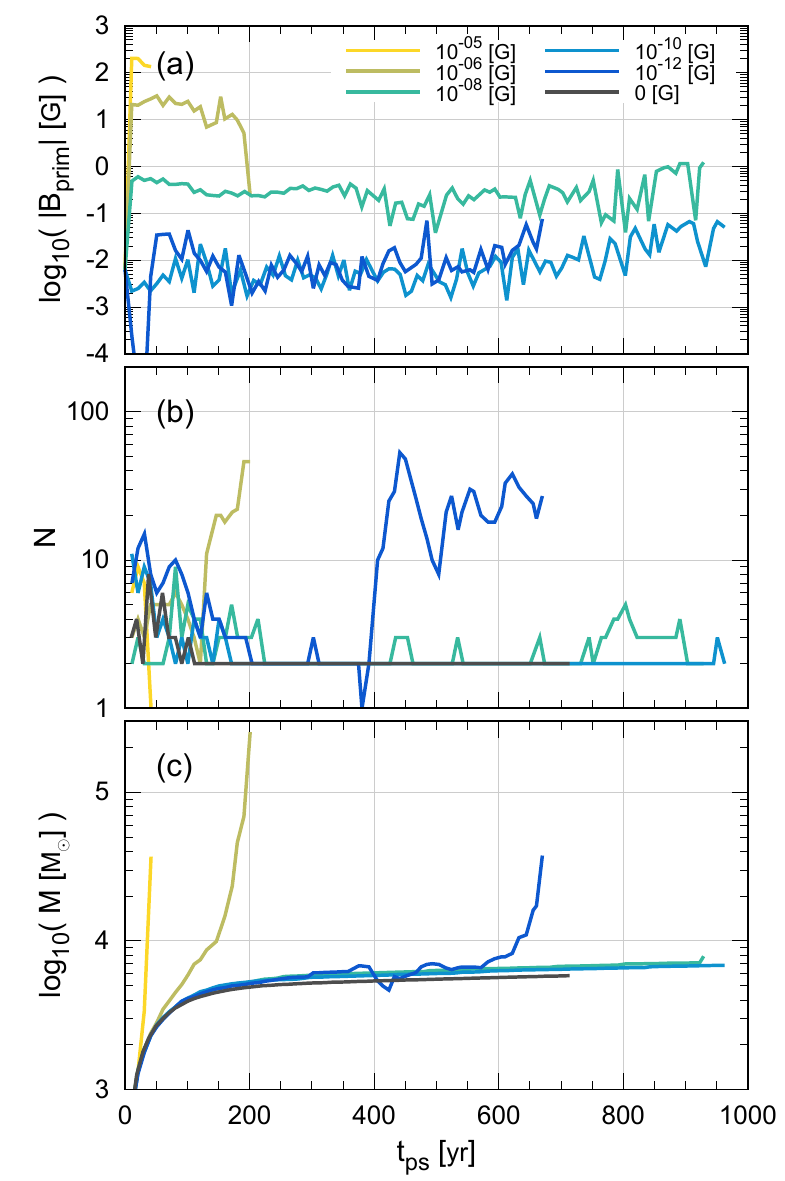}
\end{tabular}
\end{center}
\caption{
Same as Figure~\ref{f7} but for models in series M1 ($\beta_{\rm rot,0} = 10^{-1}$, left) and M3 ($\beta_{\rm rot,0} = 10^{-3}$, right).
}
\label{f9}
\end{figure*}

We consider the dependence on the initial rotation degree of the gas cloud, $\beta_{\rm rot,0}$.
We simulated models with three different $\beta_{\rm rot,0}$ (Table~\ref{t1}): $10^{-1}$ (M1 series), $10^{-2}$ (M2), and $10^{-3}$ (M3).
Figure~\ref{f8} shows the final snapshots with $B_{z,0} = 0$ and $10^{-12}$\,G models.
In cases both with and without the magnetic effects, we find that the smaller the initial rotation speed the more compact the system appears.

Figure~\ref{f9} shows the time evolution of the core's properties.
In the models with $B_{z,0} \ge 10^{-6}$\,G, the magnetic field is already strong during the collapse phase and the mass growth of the primary core speeds up with decreasing $\beta_{\rm rot,0}$.
For these models, the magnetic field strength is sufficiently strong at the end of the collapse phase and no further magnetic field amplification due to rotation is needed in the accretion phase.
The mass accretion rate increases rapidly in the lower rotation model.
On the other hand, in the models with $B_{z,0} \le 10^{-6}$\,G, the opposite dependence is confirmed.
For these models, fragmentation to make dense cores is suppressed in the slow rotation models (M3) and the magnetic field amplification during the accretion phase is also suppressed.
Thus the mass growth histories of the primary core becomes similar to the non-magnetic models (Figure~\ref{f9}c).
In the rapid rotation models (M1), the primary core mass grows faster than in the slower rotation models (M2 and M3).

In summary, for the formation of supermassive stars, the slow rotation favors the extremely strong initial magnetic field while the fast rotation favors the weak initial magnetic field.

\section{Discussion} \label{sec:dis}

\subsection{Number of dense cores}

Our simulations reveal that an enhanced mass accretion to the dense central region leads to violent fragmentation into dense cores and intermittent mergers due to the efficient angular momentum transport in the magnetized AC gas cloud.
We stopped our simulations about $1000$\,yr after the first dense core formation.
It is interesting to note that this phenomenon of fragmentation and mergers should continue even after this time because the magnetic field continues to be amplified as long as the dense cores continue to form.
The magnetic  field is amplified by the random motion of the cores.
In other words, the motion of the cores plays a role like that of a turbulent dynamo.

\begin{figure}[t]
\begin{center}
\includegraphics[width=1.00\columnwidth]{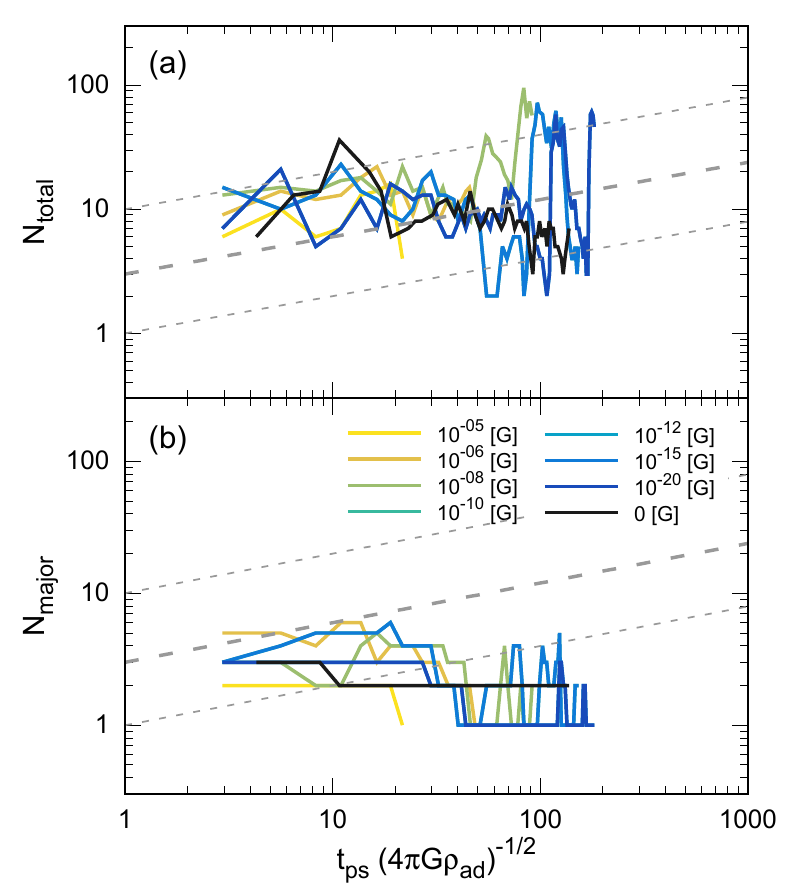}
\end{center}
\caption{
Time evolution of the core number plotted over Susa's relation \citep[figure~10 in][]{susa19}. Top panel: total core number. Bottom panel: major core number.
The dotted lines denote $3$ and $0.3$ times the dashed line.
}
\label{f10}
\end{figure}

A similar trend of the intermittent fragmentation and mergers was found for the non-magnetized zero-metallicity star formation process \citep{hirano17}.
\cite{susa19} modeled the evolution of the number of fragments with a simple phenomenological equation, roughly proportional to $t_{\rm ps}^{0.3}$, where $t_{\rm ps}$ is the elapsed time since the formation of the first protostar.
He found that most of the published numerical studies agree with the relation.
Figure~\ref{f10} overplots our results (M2 series) on the relation.
Despite the fact that our simulations initialize from a massive AC halo and treat the magnetic effects, our results do not significantly differ from this relation over a certain range (Figure~\ref{f10}a), but also cannot confirm it.
The formation of a number of dense cores due to fragmentation may continue at later times, as the strong magnetic field region is gradually expanding outward and the mass accretion rate continues to increase. 
However, we need further time integration to clearly show the time evolution of the number of cores (or fragments).

Figure~\ref{f10}b plots the number of major cores, which make up the bulk of the mass  and find that the number of the major cores becomes $\sim\!1$--$2$ for all models. 
A large number of fragments that are a significant part of the total mass are quickly incorporated into the primary core after formation.

\subsection{Cautions}

To study the magnetic effects on the accretion phase of supermassive star formation, the simulations have been simplified and the results of the current study have certain limitations.

First, we adopted the stiff EOS method as done in \citet{chon20} to compute long-term evolution of the accretion phase by restricting the numerical resolution.
For this reason, it should be noted that the dense core with $n > n_{\rm ad}$ is not strictly a star, but a region that can host a star inside.
Regarding this concern, we confirmed that the size of the dense core is close to the radius of a star of the same mass \citep{hosokawa12}, which indicates that the assumption is not significantly wrong.
There are two possible effects of the stiff EOS method on the calculation results: (1) the mass of a dense core ($M_{\rm frag}$) can be interpreted as an upper limit to the mass of the star that form inside it and (2) the artificial size of the dense core might affect the hydrodynamics of the surrounding gas.

Second, the amplification rate of magnetic field strength during the collapse phase of the star formation process is still unknown due to the dependence of the small-scale dynamo process on the numerical simulation.
Even in our simulations, it is possible that the magnetic field strength before the first dense core formation increases if we increase the numerical resolution \citep[e.g.,][]{sur10}.
However, we can infer the effect of a highly amplified magnetic field by referring to the models with greater initial magnetic field strength.
The present results show the minimum influence of the magnetic effects, and confirms that it promotes the formation of supermassive stars.

Third, we ignored turbulence, which may affect the evolution of the AC gas cloud. 
Some past studies imply the existence of turbulence in the pristine star-forming clouds   \citep{abel02,wise07,oshea07,greif08,turk09}. 
The gas accretion onto the star-forming cloud (or AC gas cloud) may drive internal turbulent motions \citep{klessen10}.
If turbulence exists in the AC gas cloud, weak magnetic fields may be amplified by the small-scale dynamo \citep{sur10,schleicher10,federrath12}. 
In such a case, the amplified magnetic field would affect the evolution of the AC gas cloud, as seen in \citet{turk12}. 
Recently, \citet{higashi21} investigated the amplification of turbulence in collapsing clouds and showed that turbulence can play an important role for the star-formation process if only weak turbulence exists before the gravitational contraction begins. 
In this study, we showed that the magnetic field is exponentially amplified around the dense core after the first protostar formation even if the initial AC gas cloud has no turbulence. 
Thus, in any case (with or without seed turbulence in the initial AC gas cloud), the magnetic field can be amplified and affect the star formation process in the AC gas cloud.

\section{Conclusion} \label{sec:con}

We have performed a set of 3D MHD simulations of a magnetized atomic-cooling gas cloud while changing two parameters: $B_{z,0}$ and $\beta_{\rm rot,0}$.
The tiny seed magnetic field can be amplified during the early accretion phase through the random motion of cores, independent of the amplification during the collapse phase.
The angular momentum transfer due to the magnetic effects increases the mass accretion rate and the coalescence rate of the fragmented dense cores.
The number of fragments is almost independent of the initial magnetic field strength. 
The accretion rate to the dense gas in the center increases as the initial magnetic field strength increases.
This causes a higher fragmentation rate but there is also a higher coalescence rate that works against a higher fragment number.

The magnetic effects can reduce the critical conditions of the supermassive star formation for each scenario, e.g., critical halo mass (atomic-cooling halo), critical intensity of the Lyman-Werner radiation (H$_2$ photo-dissociation), and critical baryon-dark matter relative velocity (streaming velocity).
This simultaneously implies that we need to revisit the formation rate of seed BHs in the early universe.
Specifically, the possibility of supermassive star formation in a  metal-enriched atomic-cooling gas cloud is an open question.

\acknowledgments
We appreciate Kazu Omukai for giving us the data of thermal evolution for primordial cloud.
We have benefited greatly from discussions with Hajime Susa.
We also thank our anonymous referee for constructive comments on this study.
This work used the computational resources of the HPCI system provided by the supercomputer system SX-ACE at Cyber Sciencecenter, Tohoku University and Cybermedia Center, Osaka University through the HPCI System Research Project (Project ID: hp190035 and hp200004), and Earth Simulator at JAMSTEC provided by 2020 Koubo Kadai.
This work was supported (in part) by JSPS KAKENHI Grant Nunbers 18J01296, 21K13960, and 21H01123 and QR (Qdai-jump Research Program) 02217 to S.H. and 17K05387, 17H02869, 17H06360, 17KK0096, 21K03617, and 21H00046 to M.N.M, and a University Research Support Grant from the National Astronomical Observatory of Japan (NAOJ).
S.B. was supported by a Discovery Grant from NSERC.



\bibliography{ms_ref}{}

\begin{thebibliography}{}
\expandafter\ifx\csname natexlab\endcsname\relax\def\natexlab#1{#1}\fi
\providecommand{\url}[1]{\href{#1}{#1}}
\providecommand{\dodoi}[1]{doi:~\href{http://doi.org/#1}{\nolinkurl{#1}}}
\providecommand{\doeprint}[1]{\href{http://ascl.net/#1}{\nolinkurl{http://ascl.net/#1}}}
\providecommand{\doarXiv}[1]{\href{https://arxiv.org/abs/#1}{\nolinkurl{https://arxiv.org/abs/#1}}}

\bibitem[{{Abel} {et~al.}(2002){Abel}, {Bryan}, \& {Norman}}]{abel02}
{Abel}, T., {Bryan}, G.~L., \& {Norman}, M.~L. 2002, Science, 295, 93,
  \dodoi{10.1126/science.295.5552.93}

\bibitem[{{Agarwal} {et~al.}(2012){Agarwal}, {Khochfar}, {Johnson}, {Neistein},
  {Dalla Vecchia}, \& {Livio}}]{agarwal12}
{Agarwal}, B., {Khochfar}, S., {Johnson}, J.~L., {et~al.} 2012, \mnras, 425,
  2854, \dodoi{10.1111/j.1365-2966.2012.21651.x}

\bibitem[{{Chatzopoulos} \& {Wheeler}(2012)}]{chatzopoulos12}
{Chatzopoulos}, E., \& {Wheeler}, J.~C. 2012, \apj, 748, 42,
  \dodoi{10.1088/0004-637X/748/1/42}

\bibitem[{{Chon} \& {Omukai}(2020)}]{chon20}
{Chon}, S., \& {Omukai}, K. 2020, \mnras, 494, 2851,
  \dodoi{10.1093/mnras/staa863}

\bibitem[{{Federrath} \& {Klessen}(2012)}]{federrath12}
{Federrath}, C., \& {Klessen}, R.~S. 2012, \apj, 761, 156,
  \dodoi{10.1088/0004-637X/761/2/156}

\bibitem[{{Greif} {et~al.}(2008){Greif}, {Johnson}, {Klessen}, \&
  {Bromm}}]{greif08}
{Greif}, T.~H., {Johnson}, J.~L., {Klessen}, R.~S., \& {Bromm}, V. 2008,
  \mnras, 387, 1021, \dodoi{10.1111/j.1365-2966.2008.13326.x}

\bibitem[{{Grete} {et~al.}(2019){Grete}, {Latif}, {Schleicher}, \&
  {Schmidt}}]{grete19}
{Grete}, P., {Latif}, M.~A., {Schleicher}, D.~R.~G., \& {Schmidt}, W. 2019,
  \mnras, 487, 4525, \dodoi{10.1093/mnras/stz1568}

\bibitem[{{Higashi} {et~al.}(2021){Higashi}, {Susa}, \& {Chiaki}}]{higashi21}
{Higashi}, S., {Susa}, H., \& {Chiaki}, G. 2021, arXiv e-prints,
  arXiv:2105.07701.
\newblock \doarXiv{2105.07701}

\bibitem[{{Hirano} \& {Bromm}(2017)}]{hirano17}
{Hirano}, S., \& {Bromm}, V. 2017, \mnras, 470, 898,
  \dodoi{10.1093/mnras/stx1220}

\bibitem[{{Hirano} \& {Bromm}(2018)}]{hirano18}
---. 2018, \mnras, 476, 3964, \dodoi{10.1093/mnras/sty487}

\bibitem[{{Hirano} {et~al.}(2017){Hirano}, {Hosokawa}, {Yoshida}, \&
  {Kuiper}}]{hirano17sv}
{Hirano}, S., {Hosokawa}, T., {Yoshida}, N., \& {Kuiper}, R. 2017, Science,
  357, 1375, \dodoi{10.1126/science.aai9119}

\bibitem[{{Hosokawa} {et~al.}(2012){Hosokawa}, {Omukai}, \&
  {Yorke}}]{hosokawa12}
{Hosokawa}, T., {Omukai}, K., \& {Yorke}, H.~W. 2012, \apj, 756, 93,
  \dodoi{10.1088/0004-637X/756/1/93}

\bibitem[{{Inayoshi} {et~al.}(2019){Inayoshi}, {Visbal}, \&
  {Haiman}}]{inayoshi19}
{Inayoshi}, K., {Visbal}, E., \& {Haiman}, Z. 2019, arXiv e-prints,
  arXiv:1911.05791.
\newblock \doarXiv{1911.05791}

\bibitem[{{Inayoshi} {et~al.}(2015){Inayoshi}, {Visbal}, \&
  {Kashiyama}}]{inayoshi15}
{Inayoshi}, K., {Visbal}, E., \& {Kashiyama}, K. 2015, \mnras, 453, 1692,
  \dodoi{10.1093/mnras/stv1654}

\bibitem[{{Klessen} \& {Hennebelle}(2010)}]{klessen10}
{Klessen}, R.~S., \& {Hennebelle}, P. 2010, \aap, 520, A17,
  \dodoi{10.1051/0004-6361/200913780}

\bibitem[{{Latif} {et~al.}(2013{\natexlab{a}}){Latif}, {Schleicher}, {Schmidt},
  \& {Niemeyer}}]{latif13}
{Latif}, M.~A., {Schleicher}, D.~R.~G., {Schmidt}, W., \& {Niemeyer}, J.
  2013{\natexlab{a}}, \mnras, 433, 1607, \dodoi{10.1093/mnras/stt834}

\bibitem[{{Latif} {et~al.}(2013{\natexlab{b}}){Latif}, {Schleicher}, {Schmidt},
  \& {Niemeyer}}]{latif13mag}
---. 2013{\natexlab{b}}, \mnras, 432, 668, \dodoi{10.1093/mnras/stt503}

\bibitem[{{Machida} \& {Doi}(2013)}]{machida13}
{Machida}, M.~N., \& {Doi}, K. 2013, \mnras, 435, 3283,
  \dodoi{10.1093/mnras/stt1524}

\bibitem[{{Machida} {et~al.}(2008){Machida}, {Matsumoto}, \&
  {Inutsuka}}]{machida08}
{Machida}, M.~N., {Matsumoto}, T., \& {Inutsuka}, S.-i. 2008, \apj, 685, 690,
  \dodoi{10.1086/591074}

\bibitem[{{Machida} \& {Nakamura}(2015)}]{machida15}
{Machida}, M.~N., \& {Nakamura}, T. 2015, \mnras, 448, 1405,
  \dodoi{10.1093/mnras/stu2633}

\bibitem[{{Matsumoto}(2007)}]{matsumoto07}
{Matsumoto}, T. 2007, \pasj, 59, 905, \dodoi{10.1093/pasj/59.5.905}

\bibitem[{{Omukai}(2001)}]{omukai01}
{Omukai}, K. 2001, \apj, 546, 635, \dodoi{10.1086/318296}

\bibitem[{{Omukai} {et~al.}(2008){Omukai}, {Schneider}, \& {Haiman}}]{omukai08}
{Omukai}, K., {Schneider}, R., \& {Haiman}, Z. 2008, \apj, 686, 801,
  \dodoi{10.1086/591636}

\bibitem[{{O'Shea} \& {Norman}(2007)}]{oshea07}
{O'Shea}, B.~W., \& {Norman}, M.~L. 2007, \apj, 654, 66, \dodoi{10.1086/509250}

\bibitem[{{Schleicher} {et~al.}(2010){Schleicher}, {Banerjee}, {Sur},
  {Arshakian}, {Klessen}, {Beck}, \& {Spaans}}]{schleicher10}
{Schleicher}, D.~R.~G., {Banerjee}, R., {Sur}, S., {et~al.} 2010, \aap, 522,
  A115, \dodoi{10.1051/0004-6361/201015184}

\bibitem[{{Schober} {et~al.}(2012){Schober}, {Schleicher}, {Federrath},
  {Glover}, {Klessen}, \& {Banerjee}}]{schober12}
{Schober}, J., {Schleicher}, D., {Federrath}, C., {et~al.} 2012, \apj, 754, 99,
  \dodoi{10.1088/0004-637X/754/2/99}

\bibitem[{{Sharda} {et~al.}(2020){Sharda}, {Federrath}, \&
  {Krumholz}}]{sharda20}
{Sharda}, P., {Federrath}, C., \& {Krumholz}, M.~R. 2020, arXiv e-prints,
  arXiv:2002.11502.
\newblock \doarXiv{2002.11502}

\bibitem[{{Sur} {et~al.}(2010){Sur}, {Schleicher}, {Banerjee}, {Federrath}, \&
  {Klessen}}]{sur10}
{Sur}, S., {Schleicher}, D.~R.~G., {Banerjee}, R., {Federrath}, C., \&
  {Klessen}, R.~S. 2010, \apjl, 721, L134, \dodoi{10.1088/2041-8205/721/2/L134}

\bibitem[{{Susa}(2019)}]{susa19}
{Susa}, H. 2019, \apj, 877, 99, \dodoi{10.3847/1538-4357/ab1b6f}

\bibitem[{{Tanaka} \& {Li}(2014)}]{tanaka14}
{Tanaka}, T.~L., \& {Li}, M. 2014, \mnras, 439, 1092,
  \dodoi{10.1093/mnras/stu042}

\bibitem[{{Turk} {et~al.}(2009){Turk}, {Abel}, \& {O'Shea}}]{turk09}
{Turk}, M.~J., {Abel}, T., \& {O'Shea}, B. 2009, Science, 325, 601,
  \dodoi{10.1126/science.1173540}

\bibitem[{{Turk} {et~al.}(2012){Turk}, {Oishi}, {Abel}, \& {Bryan}}]{turk12}
{Turk}, M.~J., {Oishi}, J.~S., {Abel}, T., \& {Bryan}, G.~L. 2012, \apj, 745,
  154, \dodoi{10.1088/0004-637X/745/2/154}

\bibitem[{{Umeda} {et~al.}(2016){Umeda}, {Hosokawa}, {Omukai}, \&
  {Yoshida}}]{umeda16}
{Umeda}, H., {Hosokawa}, T., {Omukai}, K., \& {Yoshida}, N. 2016, \apjl, 830,
  L34, \dodoi{10.3847/2041-8205/830/2/L34}

\bibitem[{{Wise} \& {Abel}(2007)}]{wise07}
{Wise}, J.~H., \& {Abel}, T. 2007, \apj, 671, 1559, \dodoi{10.1086/522876}

\bibitem[{{Wise} {et~al.}(2019){Wise}, {Regan}, {O'Shea}, {Norman}, {Downes},
  \& {Xu}}]{wise19}
{Wise}, J.~H., {Regan}, J.~A., {O'Shea}, B.~W., {et~al.} 2019, \nat, 566, 85,
  \dodoi{10.1038/s41586-019-0873-4}

\bibitem[{{Woods} {et~al.}(2019){Woods}, {Agarwal}, {Bromm}, {Bunker}, {Chen},
  {Chon}, {Ferrara}, {Glover}, {Haemmerl{\'e}}, {Haiman}, {Hartwig}, {Heger},
  {Hirano}, {Hosokawa}, {Inayoshi}, {Klessen}, {Kobayashi}, {Koliopanos},
  {Latif}, {Li}, {Mayer}, {Mezcua}, {Natarajan}, {Pacucci}, {Rees}, {Regan},
  {Sakurai}, {Salvadori}, {Schneider}, {Surace}, {Tanaka}, {Whalen}, \&
  {Yoshida}}]{woods19}
{Woods}, T.~E., {Agarwal}, B., {Bromm}, V., {et~al.} 2019, \pasa, 36, e027,
  \dodoi{10.1017/pasa.2019.14}

\bibitem[{{Xu} {et~al.}(2008){Xu}, {O'Shea}, {Collins}, {Norman}, {Li}, \&
  {Li}}]{xu08}
{Xu}, H., {O'Shea}, B.~W., {Collins}, D.~C., {et~al.} 2008, \apjl, 688, L57,
  \dodoi{10.1086/595617}

\bibitem[{{Yang} {et~al.}(2020){Yang}, {Wang}, {Fan}, {Hennawi}, {Davies},
  {Yue}, {Banados}, {Wu}, {Venemans}, {Barth}, {Bian}, {Boutsia}, {Decarli},
  {Farina}, {Green}, {Jiang}, {Li}, {Mazzucchelli}, \& {Walter}}]{yang20}
{Yang}, J., {Wang}, F., {Fan}, X., {et~al.} 2020, arXiv e-prints,
  arXiv:2006.13452.
\newblock \doarXiv{2006.13452}

\bibitem[{{Yoon} {et~al.}(2012){Yoon}, {Dierks}, \& {Langer}}]{yoon12}
{Yoon}, S.~C., {Dierks}, A., \& {Langer}, N. 2012, \aap, 542, A113,
  \dodoi{10.1051/0004-6361/201117769}

\end{thebibliography}
\bibliographystyle{aasjournal}

\end{document}